\documentclass[a4paper,fleqn,usenatbib]{mnras}

\usepackage{mathptmx}

\usepackage[T1]{fontenc}
\usepackage{ae,aecompl}

\usepackage{graphicx}	
\usepackage{amsmath}	
\usepackage{amssymb}	

\title[Untangling galaxy components]{Untangling galaxy components:
  full spectral bulge--disc decomposition}
\author[M. Tabor et al.]{
Martha Tabor,$^{1}$\thanks{E-mail: martha.tabor@nottingham.ac.uk}
Michael Merrifield,$^{1}$
Alfonso Arag\'{o}n-Salamanca,$^{1}$
\newauthor
Michele Cappellari,$^{2}$ Steven P. Bamford$^{1}$and Evelyn Johnston$^{3}$
\\ \\
$^{1}$School of Physics and Astronomy, University of Nottingham, University Park, Nottingham, NG7 2RD, UK\\
$^{2}$Sub-department of Astrophysics, Department of Physics, University of Oxford, 
Denys Wilkinson Building, Keble Road, Oxford OX1 3RH, UK\\
$^{3}$European Southern Observatory, Alonso de Cordova 3107, Vitacura, Santiago, Chile
}

\date{Accepted XXX. Received YYY; in original form ZZZ}

\pubyear{2015}

\begin{document}
\label{firstpage}
\pagerange{\pageref{firstpage}--\pageref{lastpage}}
\maketitle

\begin{abstract}

To ascertain whether photometric decompositions of galaxies into bulges and disks are 
astrophysically meaningful, we have developed a new technique to decompose spectral data 
cubes into separate bulge and disk components, subject only to the constraint that they 
reproduce the conventional photometric decomposition.  These decompositions allow us to 
study the kinematic and stellar population properties of the individual components and 
how they vary with position, in order to assess their plausibility as discrete elements, 
and to start to reconstruct their distinct formation histories. An initial application 
of this method to CALIFA integral field unit observations of three isolated S0 galaxies 
confirms that in regions where both bulge and disc contribute significantly to the flux 
they can be physically and robustly decomposed into a rotating dispersion-dominated 
bulge component, and a rotating low-dispersion disc component. Analysis of the 
resulting stellar populations shows that the bulges of these galaxies have a range of ages relative
to their discs, indicating that a variety of processes are necessary to describe their evolution.
This simple test case indicates the broad potential 
for extracting from spectral data cubes the full spectral data of a wide variety of individual 
galaxy components, and for using such decompositions to understand the interplay between these 
various structures, and hence how such systems formed.

\end{abstract}

\begin{keywords}
galaxies: kinematics and dynamics -- galaxies: elliptical and lenticular 
\end{keywords}



\section{Introduction}
Almost since the study of galaxy evolution first began, bulge--disc
decompositions have been used to try and understand the separate
formation and evolution histories of each component. Originally, many
galaxies were thought to comprise a central spheroidal bulge with a de
Vaucouleurs law surface brightness profile \citep{DeVaucouleurs1948}
combined with a more extended exponential disc \citep{freeman1970}.
This division fits nicely with a dynamical picture in which part of
the system consists of stars on largely random orbits, resulting in a
spheroidal component with little net rotation, while the remainder is
made up of stars in well-ordered rotation, producing a
highly-flattened disc.  Thus, bulge--disc decomposition offered a neat
and convenient way of separating the different parts of a galaxy, and
hence exploring the formation histories of these dynamically-distinct
features.  To this end, increasingly sophisticated software such as
GALFIT \citep{Peng2002}, GIM2D \citep{simard1998} and BUDDA \citep{deSouza2004} has been developed to formalise and automate
the process of fitting such multiple components.

However, as the level of detail in observations has increased,
evidence has appeared indicating that bulges and discs are not as
simple and distinct as was first thought. With the high image
resolution made possible with HST, it was found that many bulges
display disc-like features such as spiral structure, inner discs and
bars, and often follow shallower surface brightness profiles than
originally thought, with S\'{e}rsic indices as low as the value of
unity that is characteristic of an exponential disc \citep{Sersic1963,
  Andredakis1994, Andredakis1995, Courteau1996, Carollo1998,
  MacArthur2004}.  In addition, work on the structure of bulges with
boxier shapes has found that these components are actually thickened
bars, an element that had originally been associated with galactic
discs \citep{Kuijken1995, Bureau1999}.  Such developments have blurred
the distinction between discs and bulges, and thus raised the question
of whether the components obtained by decomposing a galaxy
photometrically really represent individual stellar populations with
distinct kinematics, or whether these decompositions are just a
convenient way of parameterising galaxy light distributions.

The key to addressing this question is provided by spectra: since
their absorption lines contain information on both stellar kinematics
and populations, analysis of spectral data can in principle be used to
test whether the individual bulge and disc components form coherent
distinct entities in terms of their dynamics and stellar properties,
and hence whether the decomposition makes any intrinsic physical
sense.

Historically, spectral observations were obtained using long-slit
spectrographs, which provided a one-dimensional cut through the light
distribution of a galaxy. \citet{Johnston2012} developed a
technique to decompose these profiles into bulge and disc components
at each wavelength provided by the spectrograph, thus producing a
direct measure of the amount of bulge and disc light as a function of
wavelength; in other words, it extracted separate bulge and disc
spectra, which would allow the stellar populations in these components
to be compared.  More recently, they have generalised the technique to
analyse integral-field unit (IFU) data, which produce spectra right
across the face of a galaxy, and hence generate a three-dimensional
data cube comprising a spectrum at each location on the sky. By
applying GALFITM \citep{2013MNRAS.430..330H}, a modified version of GALFIT, 
to an ``image slice'' through the data cube at each
wavelength, it is possible to calculate the relative contributions of
bulge and disc at that wavelength, and hence calculate a separate
spectrum for each component as before, but now based on a more
reliable two-dimensional image rather than just a one-dimensional cut
(Johnston et al., in prep).

Although this approach has yielded new insights into the formation
sequence of bulges and discs in S0 galaxies \citep{Johnston2014}, it
is not well suited to addressing the question of the distinctness of
the components.  Because it reduces each component to a single
spectrum, it loses most of the information relating to variations with
position in the galaxy, which would tell us how homogeneous each
component is.  In addition, it throws away all the kinematic
information in the spectra, even though some of the clearest defining
features of bulges and discs are related to their distinct dynamical
properties [see \citealt{Cappellari2016} for a review].

The presence of multiple galaxy components has also been previously explored through a 
purely kinematic approach.
\citet{Bender1990} found that the asymmetric shape of the line-of-sight velocity distribution 
along the major axis of discy elliptical galaxy NGC~4621 was best 
explained by the presence of two gaussians with different mean velocities and velocity dispersions, i.e. a bulge and a
disc component. Subsequent studies such as \citet{Scorza1995} used this effect to explore the kinematics of 
bulge and disc components for a wider sample of discy ellipticals. 
However, this approach is limited in terms of the spatial information, looking only at the behaviour along 
major axis of the galaxy, and does not provide information on the stellar populations of the 
components, which is also a key element in determining whether the components are truly distinct.

In this paper, we therefore attempt a new approach to decomposing
galaxies into bulge and disc components, which retains both the kinematic
and stellar population information, and allows freedom for the properties of each component to
vary with position, subject only to the global constraint that
the relative fluxes of bulge and disc component at each location match
those produced by image decomposition.  Thus these models, by
construction, reproduce the bulge and disc components of the
conventional image fitting technique, but also fit the full spectral
data cube for the galaxy.  By investigating the stellar kinematics and
populations of the resulting fitted components, and how they vary with
position, it will be possible to test whether the imposed image
decomposition produces coherent underlying physical sub-systems.

\section{Test data}\label{sec:data}

To develop this approach, we need a few representative galaxy observations to
test that the method is viable with the quality of data currently
being obtained.  There are an expanding number of IFU galaxy surveys
from which such a test case could be drawn, including SAURON
\citep{deZeeuw2003}, ATLAS$^{\rm 3D}$ \citep{Cappellari2011}, SAMI
\citep{Bryant2015} and MaNGA \citep{Bundy2015}.  For this analysis, we
have chosen to use data from the recently-completed Calar Alto
Integral Field Area (CALIFA) survey \citep{Sanchez2012}, which obtained spatially-resolved
spectroscopy of 600 nearby ($0.005 \leq z \leq 0.03$) galaxies over a
wide range of morphologies.  We analyse an observation in the V1200
setup, which yielded spectra with a wavelength range of  3650--4840\AA, 
and a spectral resolution of 2.3\AA\ FWHM, corresponding to a velocity
resolution of $\sim80$ km s$^{-1}$ at the blue end and $\sim60$ km s$^{-1}$ at the red end of
the spectrum.  A full description of the data can be be
found in \citet{Garcia2015}.

\begin{figure*}
\begin{center}
	\includegraphics[width=0.8\textwidth]{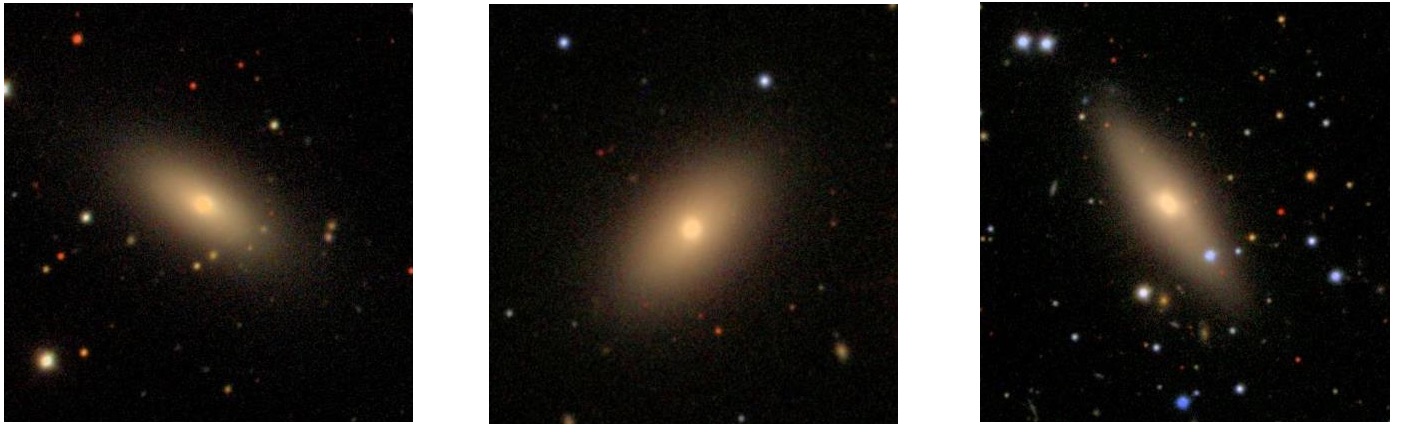} 
        \caption{SDSS images of the three test galaxies over the CALIFA field of view; from left to right NGC~528, NGC~7671 and NGC~6427.\label{fig:gals}}
 \end{center}
 \end{figure*}

From the CALIFA sample, we have selected three field S0 galaxies: 
NGC~528, NGC~7671 and NGC~6427. The SDSS colour images of these
galaxies are shown in Figure~\ref{fig:gals},
and the galaxies properties are summarised in Table~\ref{tab:galpar}. 
With their simple structure, all 
containing clear bulge and disc components and no visible dust, 
spiral structure or bar, they make good first test examples
of such a bulge--disc decomposition method. Selecting this
Hubble type and environment also allows us to compare and contrast the result with the
previous analysis by \citet{Johnston2014} in their study of cluster S0s. 

In order to fully determine the strengths and limitations of the method, we have chosen 
to explore the decomposition of one galaxy, NGC~528, in particular detail. At an inclination of
$\sim 70$~degrees, it clearly displays both bulge and disc, while
still ensuring that enough of any rotational motion is projected into
the line of sight to be detected. It was also chosen for its
brightness: it has a K-band absolute magnitude of $M_K \approx -24.36$ \citep{Skrutskie2006};
using eq.(2) of \citet{Cappellari2013} this luminosity can be converted into a large stellar 
mass of $M_\ast\approx1.5 \times 10^{11}M_\odot$, 
ensuring
that any kinematic structure will be well resolved with the CALIFA
spectral resolution. The following sections will outline the method as applied
to this galaxy.

\begin{table}
\begin{center}
\begin{tabular}{ c|cccc }
Galaxy & Kmag & $M/M_\odot$ & I($^{\circ}$) & Distance modulus\\
\hline
NGC~528 & $-24.36$ & $1.5\times10^{11}$ & 70 & $34.24$\\
NGC~7671 & $-24.05$ & $5.9\times10^{10}$ & 75 & $33.49$\\
NGC~6427 & $-23.43$ & $4.0\times10^{10}$ & 90 &  $33.91$\\
\hline
\end{tabular}
\caption{Absolute magnitude, stellar mass (converted from K-band magnitude using eq.(2) of \citealt{Cappellari2013}), inclination and 
redshift distance modulus \citep{Makarov2014} 
of galaxy sample.\label{tab:galpar}}
\end{center}
\end{table}

\section{Photometric bulge--disc decomposition}\label{sec:phot}

To perform the initial bulge--disc decomposition that tells us how
much light to ascribe to each component at each point in the galaxy,
we have first collapsed the data cube in the spectral
direction to produce a fairly crude image of the galaxy at the
appropriate spatial resolution.  We applied GALFIT to this image to
fit an exponential disc plus a spheroidal component with a S\'{e}rsic
profile for which the index was left as a free parameter.  We also
left the position angle and flattening of each component as free
parameters.  

The specific point-spread function (PSF) for this observation is not
part of the currently-released CALIFA data, so we modelled it as a
narrow Moffat profile plus a broad low-intensity Gaussian as described
in \citet{Garcia2015}, and varied its Moffat parameters to minimise
the residuals between model and image data.  The best fit was found to
result from a PSF with a FWHM of $2.2\,{\rm arcsec}$, although the subsequent
results do not depend sensitively on the exact value adopted.

\begin{figure*}
\begin{center}
	\includegraphics[width=0.8\textwidth]{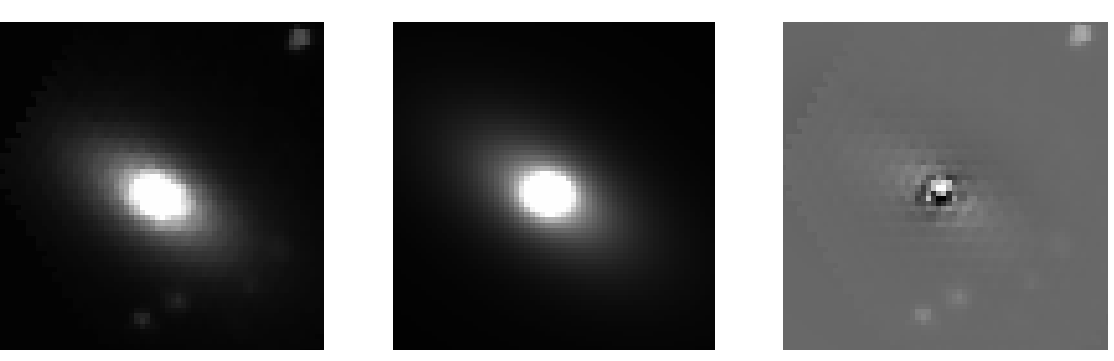} 
        \caption{The collapsed CALIFA data cube for NGC~528, the GALFIT
          best fit model, and the residuals (galaxy minus
          model) \label{fig:galf}}
 \end{center}
 \end{figure*}

The parameters of the derived best-fit galaxy models for the three sample galaxies are listed in
Table~\ref{tab:GALFIT2}, and an indication of the quality of the fit is
provided in Figure~\ref{fig:galf}, which shows data, model and
residuals for NGC~528.  The fit is generally very good, although the central
asymmetric residual indicates that the galaxy seems to have a
slightly off-centre nucleus, but this phenomenon has no effect on what
follows.  It is also interesting to note from Table~\ref{tab:GALFIT2} that this galaxy
clearly favours an almost exponential bulge with a S\'{e}rsic index
near unity, and is therefore an example of the non-classical bulge
behaviour mentioned above.

\begin{table*}
\begin{center}
\begin{tabular}{ c|cc|cc|cc }
  & \multicolumn{2}{c}{NGC~528} & \multicolumn{2}{c}{NGC~7671} & \multicolumn{2}{c}{NGC~6427} \\
Parameter & Bulge & Disc & Bulge & Disc & Bulge & Disc\\
\hline
Half light radius/Scale length & $2.21''\pm0.09''$ & $9.17''\pm0.11''$ & $1.6''\pm0.002$'' & $8.68''\pm0.05''$ & $2.90''\pm0.04''$ & $9.88''\pm0.11''$\\
Axis ratio & $0.78\pm0.01$ & $0.52\pm0.01$ & $0.93\pm0.01$ & $0.67\pm0.002$ & $0.797\pm0.003$ & $0.474\pm0.003$\\
S\'{e}rsic Index & $1.07\pm0.04$ & $1.0$ (fixed) & 0.99$\pm$0.04 & 1.0 (fixed) & 0.93$\pm$0.03 & 1.0 (fixed)\\
Position Angle & $-89.8\pm2.57$ & $59.02\pm0.15$ & 18.6$\pm$7.6 & $-43.6\pm0.3$ & $64.8\pm2.6$ & $34.7\pm0.3$\\
Flux Fraction & $0.275\pm0.005$ & $0.725\pm0.005$ &  $0.277\pm0.002$ & $0.723\pm0.002$ & $0.320\pm0.002$ & $0.680\pm0.002$\\
\hline
\end{tabular}
\caption{Parameters of GALFIT photometric decomposition of collapsed
  data cube image of NGC~528, NGC~7671 and NGC~6427. Note the errors shown 
  are statistical estimates directly from GALFIT, which underestimate the true uncertainties.\label{tab:GALFIT2}}
\end{center}
\end{table*}

\section{Spectroscopic bulge--disc decomposition}

\subsection{Method}\label{sec:meth}

In order to carry out the spectral decomposition, we need to fit bulge and disc components, weighted by their relative contribution at each 
spatial coordinate of the data cube, to the spectrum at that point. To this end, we have used the Python version of the penalised 
pixel-fitting code \citep{Cappellari2004}, \footnote{Available from \url{http://purl.org/cappellari/software}} updated as described in 
\citet{Cappellari2017} which, from version 5.1, allows multiple components with different populations and kinematics 
to be fitted to spectra simultaneously (see \citealt{Johnston2013}).    

For a given set of nonlinear parameters, pPXF solves a linear system for the weights of the spectral templates 
and additive polynomials, with the extra constraint that the weights of the spectral templates must be non-negative  
[eq. 3 in \citet{Cappellari2004}]. This linear least-squares problem is solved by pPFX as a Bounded Variables 
Least Squares problem, with the specific BVLS method by \citet{Lawson1995}.

Here we want to enforce an additional constraint to the problem: we require the template spectra describing the 
bulge component to contribute a prescribed fraction $f_{\rm bulge}$ of the total flux in the fitted spectrum
\begin{equation}
f_{\rm bulge}=\frac{\sum w_{\rm bulge}}{\sum w_{\rm bulge} + \sum w_{\rm disk}},
\end{equation}
where $w_{\rm bulge}$ and $w_{\rm disk}$ are the weights assigned to the set of spectral templates used to 
fit the bulge and disk respectively.

This constraint could be implemented by using a generic quadratic programming algorithm to solve the linear 
least-squares sub-problem, while including equation~(1) as an exact linear equality constraint. Instead, 
we chose a simpler route, which requires minimal changes to the pPXF algorithm. It consists of 
enforcing the equality constraint by adding the following single extra equation to the linear least-squares sub-problem
\begin{equation}
\frac{1}{\Delta}\left[(f_{\rm bulge}-1)\sum w_{\rm bulge} + f_{\rm bulge}\sum w_{\rm disk}\right]=0,
\end{equation}
with the parameter $\Delta$ set to a very small number (e.g. $10^{-9}$), which specifies the relative accuracy at which 
this equation needs to be satisfied. When both the spectral templates and the galaxy spectrum are normalized to have a 
mean flux of order unity, the best fitting weights $w_{\rm bulge}$ and $w_{\rm disk}$  are generally smaller than unity 
and equation~(2) is satisfied to numerical accuracy. This new option is activated by setting the keyword \textsc{fraction} 
from the Python pPXF version 5.2. Activating this option results in a negligible execution time penalty of the pPXF fit, 
when compared to an unconstrained fit.

\subsection{Tests}\label{sec:test}
The pPXF code requires that one specifies the base set of stellar
populations from which the model is constructed.  For this analysis,
we have adopted the single stellar population templates of \citet{Vazdekis2010}, providing high-resolution
spectra for stellar populations with 7 different metallicities ranging from
$-2.32$ to $0.22$ and 42 ages from $0.06\,{\rm ~Gyr}$ to $17.8\,{\rm ~Gyr}$, resulting in 
a total catalogue of 350 spectra.
An age range that extends beyond
the age of the Universe is required to account for uncertainties in
the stellar population models. Linear combinations of this extensive
library should provide a reasonable match to most stellar populations
that we are likely to encounter, and offer at least
an initial estimate of their ages and metallicities.

\begin{figure*}
\begin{center}
	\includegraphics[width=0.9\textwidth]{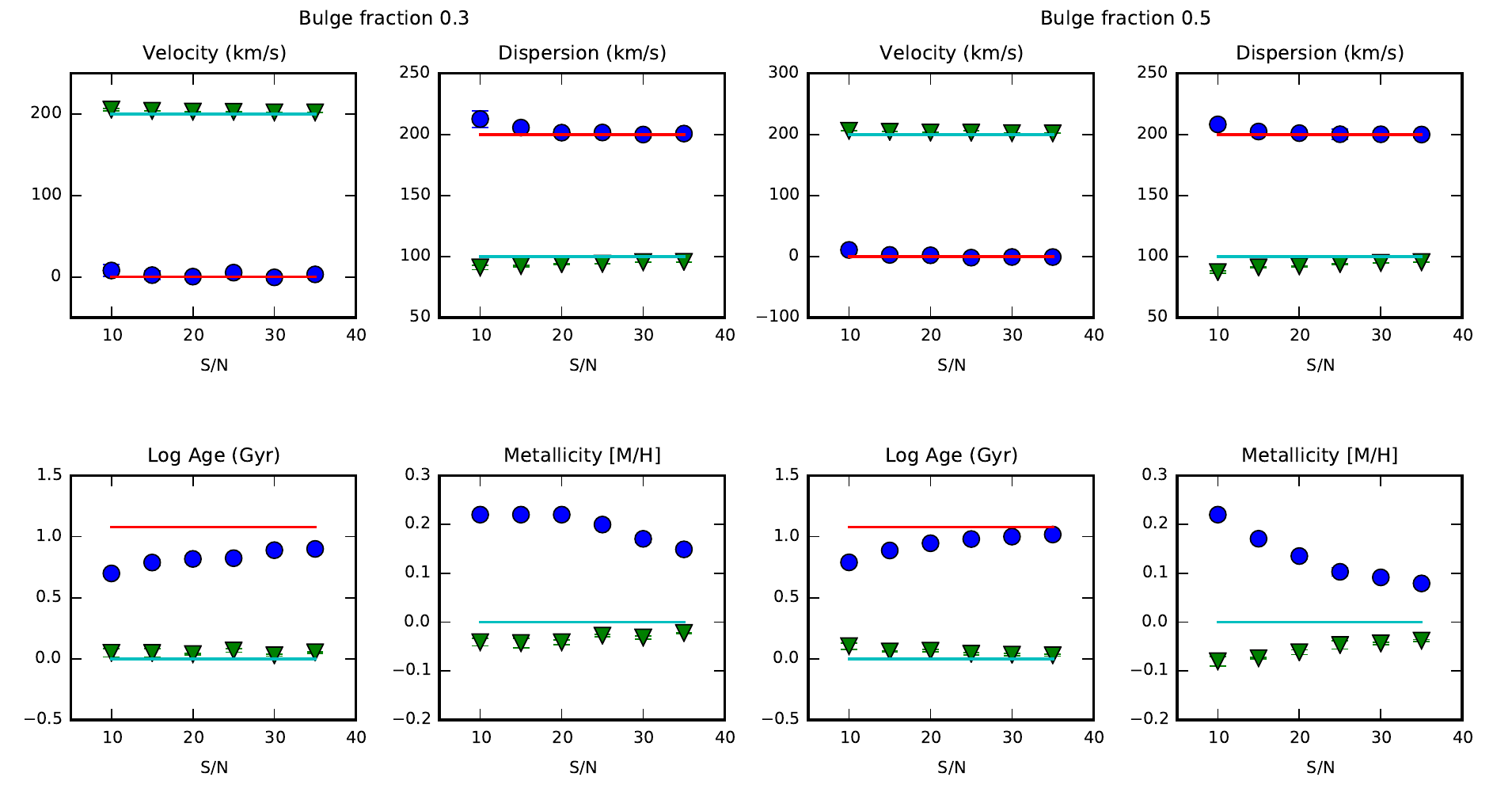} 
        \caption{Best-fit kinematics and stellar populations for a simulated galaxy spectrum with bulge fractions of
                  0.3 and 0.5. Blue circles represent
                  the median values of 10 realisations with varying initial estimates for the bulge, and green triangles 
                  represent the same for the disc. The error bars show the standard deviation of the 10 realisations. 
                  The solid line shows the actual values given to the components.
		  \label{fig:test}}
\end{center}
\end{figure*}

In general, the signal-to-noise ratio of the individual spectra in the
data cube is not sufficient to fit simultaneously for kinematics and
populations of multiple components.  In order to determine at what
signal-to-noise level this fit becomes possible, we have tested the method by
decomposing a simulated galaxy spectrum composed of two stellar populations of different
ages of 1 and 12 Gyrs, and with kinematics typical of a bulge and disc. 
To explore the full capabilities of the method we have looked at models with 
a galaxy bulge-to-total ratio of 0.3 and 0.5, and for each we vary the simulated signal-to-noise ratio
from 10 to 35.
To ensure the model is actually converging on the proper solution and does not
simply return values close to the initial estimates, we have performed ten realisations at
each signal-to-noise ratio, each time setting the initial kinematic estimates of both velocity and sigma to be a 
random value within 50 km~s$^{-1}$ of the actual value.
The best-fit kinematics and stellar populations at each signal-to-noise ratio are
shown in Figure~\ref{fig:test}.
 
This test demonstrates the effectiveness of the method, showing
that when both components
contribute a significant amount to the flux their kinematics 
can be accurately determined and are not heavily dependant on the initial kinematic estimates. 
To retain as much spatial information as possible while avoiding significant biases, particularly in the derived metallicity, 
a signal-to-noise ratio of 25 was adopted for the decompostition.
In order to obtain spectra of this quality from the CALIFA data cube, we co-added the data spatially
using the Python version of the Voronoi binning code of \citet{Cappellari2003}.

We performed an additional test on the real CALIFA data cube, taking a sample of 
bins along the major axis of the galaxy and performing the decomposition for each bin over a 
grid of bulge and disk velocities, from $-400\,$ km s$^{-1}$ to $400\,$ km s$^{-1}$. pPXF uses 
a local non-linear optimization algorithm to fit the kinematics, starting from an initial guess 
for the parameters. However, with multiple kinematic components, the $\chi^2$ will generally 
present multiple local minima and there is no guarantee pPXF will converge to the global one. 
A simple brute-force way of finding the global minimum consists of sampling a regular grid of 
starting guesses with pPXF. By constraining the velocities of the 
fit to be within
the pixel of the velocity grid, while letting the velocity dispersions vary freely, pPXF is forced to cover the full range 
of velocities, and 
we can therefore make sure we find the global rather than a possible local minimum for the $\chi^2$.

\begin{figure*}
\begin{center}
	\includegraphics[width=\textwidth]{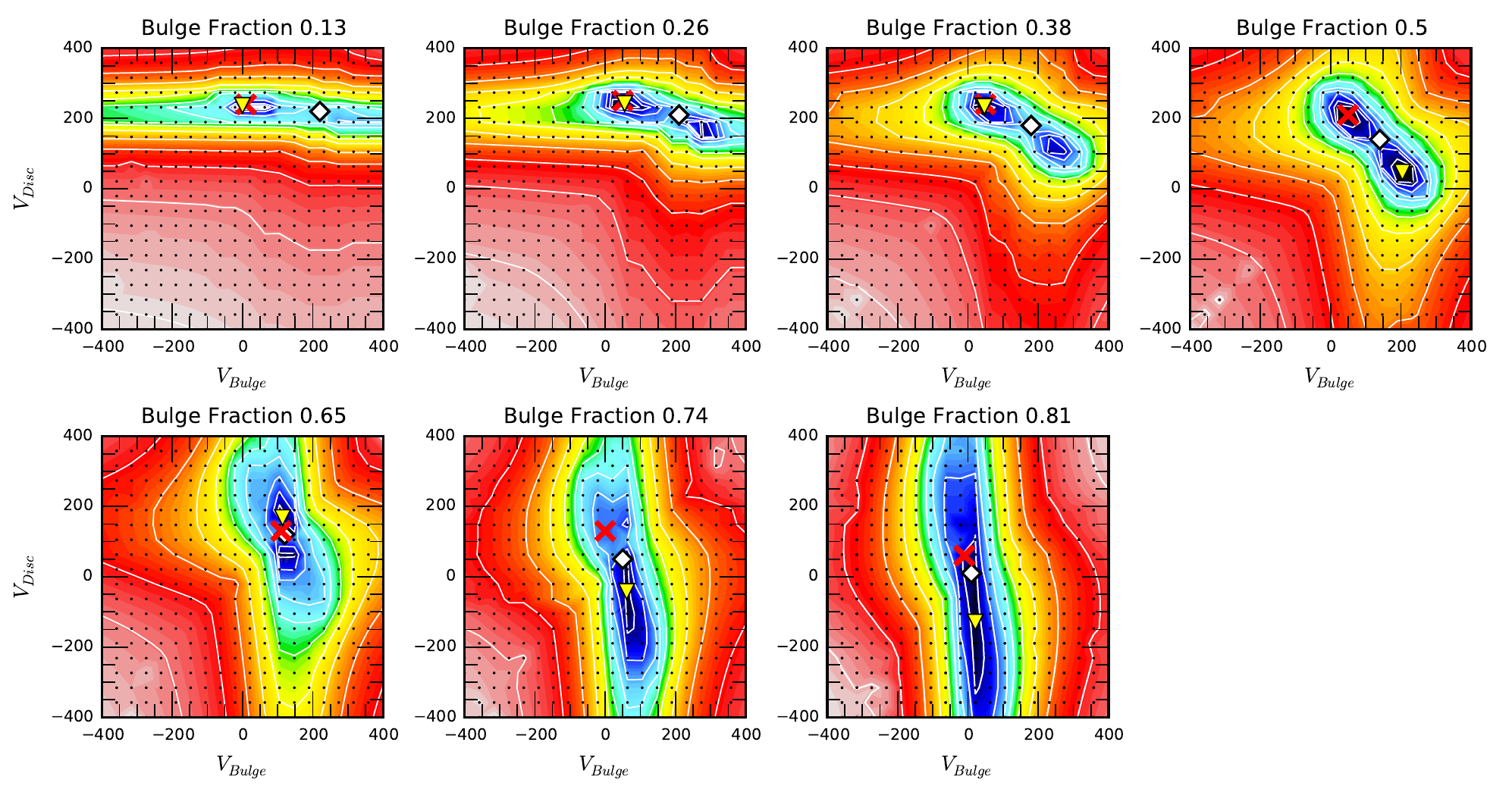} 
        \caption{Chi-squared maps for a sample of bins along the major axis of the galaxy. One sigma contours are also
                 plotted. Red crosses show the 
                 velocities found when pPXF is left free to vary, white diamonds show the velocity of a one component
                 fit, and yellow triangles show the velocities for the lowest chi-squared grid pixels. When the yellow triangles and
                 red crosses coincide, pPXF is effectively finding the global minimum from the estimated starting conditions.
		  \label{fig:chtest}}
\end{center}
\end{figure*}

The chi-squared map over this grid is shown in Figure~\ref{fig:chtest} including one sigma contours, with the 
minimum chi-squared locations indicated by the yellow triangles.
In order to evaluate how well pPXF finds this global minimum, the best-fit velocities found by pPXF when it is left 
free to vary are plotted as red crosses. As starting conditions for this free fit we took the velocity of a single 
component fit for the disc velocity, and the average velocity dispersion in the 
disc dominated region of the galaxy, here $144\,$ km s$^{-1}$ for the disc dispersion. For the bulge we estimated zero rotational
 velocity, and a velocity dispersion value of 
$200\,$ km s$^{-1}$.

This test gives us an insight both into how well pPXF finds the true best fit to the data when left free, 
and how accurately the kinematics of the components can really be determined overall
for this quality of data. 

As is to be expected, when a component 
dominates the flux pPXF effectively finds the best fit to the kinematics for this component when left 
free, and the velocity of this component 
is very well constrained, with
the one-sigma contours covering a small range of velocities. 
However, the kinematics of the component 
with less flux are less likely to be found by pPXF and tend to show some degeneracy. This is especially apparent in 
bulge-dominated regions of the galaxy,
with the one sigma contours stretching over a wide range of disc velocities. This is not surprising considering that in the very
central regions of the galaxy there will be little difference in velocity between the two components, making it almost 
impossible to separate the LOSVDs effectively.

The tests also demonstrate that in the central regions of the galaxy, when pPXF is left free it
does not always find the true chi-squared minima. Though the difference here is generally less than 3 sigma, for data of 
lower signal-to-noise and lower resolution this may not be the case. Therefore, we decided that to ensure that the true 
chi-squared minimum is found each time, when performing the decomposition on the
full galaxy a simple 5x5 grid of velocities should be used for every voronoi bin.



Also important to note is that when both components contribute comparable fluxes two clear minima in chi-squared are visible,
which have significantly lower chi-squared than the single component fit. 
This shows that the 
spectrum is better fit by two 
Gaussian components with different velocities and velocity dispersions, 
rather than one single Gaussian component. This is a first indication that two kinematically-distinct 
components do appear to be present in this galaxy.

It is worth noting that in conducting these tests we have built the machinery 
necessary to check the reliability of the decomposition process for
any additional parameters, be it signal-to-noise ratio, initial kinematic estimates,
component flux contribution etc. which will become useful when exploring new
data and more complex systems.

\subsection{NGC 528}

In order to determine the kinematics of the bulge and disk implied by the assumed 
bulge-disk photometric decomposition, we have fitted each
Voronoi bin with a kinematic model comprising two components with Gaussian
line-of-sight velocity distributions. As demonstrated above (Section~\ref{sec:test}), in order to be 
certain that the true best fit to the spectra is being found, it is safest to force pPXF to 
cover a wide range of velocities for each component. For each Voronoi bin we therefore 
take the 
best fit to the spectrum across a 5x5 grid of 
velocities ranging from $-350\,$ km s$^{-1}$ to $350\,$ km s$^{-1}$. By constraining the resulting 
velocities of each fit to be within the pixel size of the grid, pPXF is forced to cover the whole velocity
range and therefore find the true chi-squared minimum. 
The only additional constraint on the decomposition is that
one component must reproduce the
flux of the bulge derived in the photometric decomposition, while the
other component must reproduce the disc flux.  Both velocity dispersions and the stellar populations of the two
components were left as free parameters. Since a change in kinematics can be slightly offset by altering the template spectra 
selected for the fit, it was deemed safest to continue to use the whole 
catalogue of 350 spectra to fit each component.

\begin{figure*}
\begin{center}
	\includegraphics[width=0.8\textwidth]{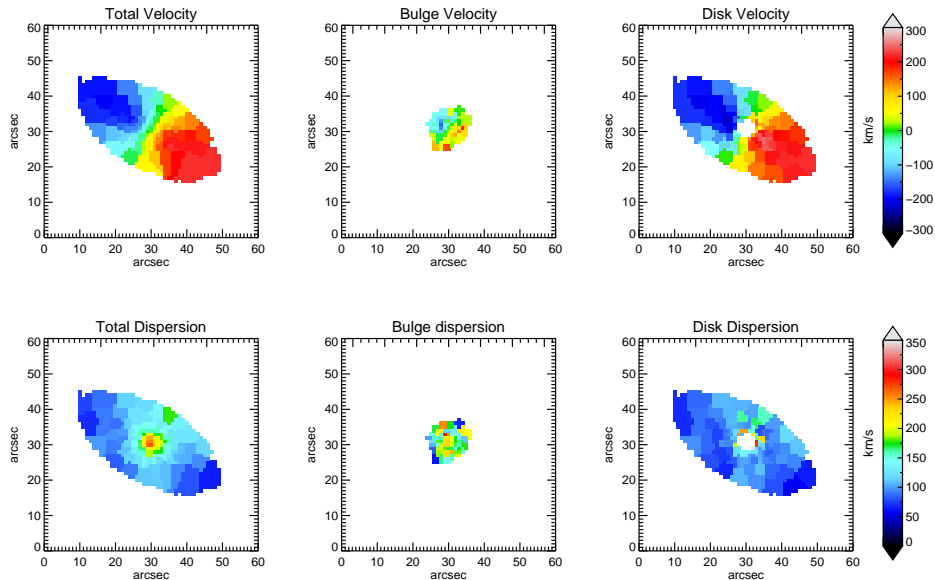} 
 \caption{Kinematic maps for a one component fit (left) and two component fit (centre and right) for NGC~528\label{fig:bulger}} 
\end{center}
\end{figure*}

The resulting kinematic maps are presented in Figure~\ref{fig:bulger}.  To
avoid plotting noise and to give a clear indication where each
component dominates, we have only reproduced the kinematic maps in
regions where a component contributes at least 30\% of the total flux. 
We found that below this level pPXF struggled to effectively distinguish 
the LOSVD of the lower flux component. The outer limit of the disc is 
determined by a minimum signal-to-noise threshold required for the 
Voronoi binning process, discarding pixels in the collapsed data-cube where 
the signal-to-noise ratio is too low.

It is apparent from this figure that this process decomposes the
galaxy into remarkably coherent individual components, with a
well-behaved symmetrically-rotating disc that has a low dispersion at
all radii, and a symmetric high-dispersion rotating bulge with a dispersion
profile that decreases in an orderly manner with radius.  Note that
the symmetric appearance of the kinematics is in no way imposed on the
results: each spatial bin is decomposed independently, so the
coherence of the map is a direct indicator of the consistency and
reproducibility of the results, even down to the level where the
component is only a 30\% contributor to the total light.

The derived velocity dispersion of the bulge is high, but quite
consistent with the central dispersion of $264\,$ km s$^{-1}$ found by
\citet{Scodeggio1998}, and the expected properties of such a massive
galaxy (see Section~\ref{sec:data}).  The mass of the galaxy is also
reflected in the rotation speed of the disc, which, after correcting
for inclination, implies a value of a little over $300\,$ km s$^{-1}$.
Astrophysically, the rapid rotation of a low-S\'{e}rsic-index bulge is not surprising and agrees
with multiple studies finding that these flatter bulges tend to display disc-like
characteristics (see \citealt{Kormendy2004} and references therein). The bulge 
rotation is also consistent with the finding that bulges and disks in fast 
rotator ETGs are best described by constant anisotropy (see \citealt{Cappellari2016} for a review).

\subsection{NGC~7671 and NGC~6427}

Two other galaxies in this initial test, NGC~7671 and NGC~6427,
were analysed in the same way.
The galaxy magnitudes, masses and inclinations are shown in Table~\ref{tab:galpar} and
the parameters of the GALFIT photometric decomposition are shown in Table~\ref{tab:GALFIT2}.
Similar to NGC~528, both galaxies also possess low-S\'{e}rsic-index bulges, allowing us to gain
further insight into these systems.

\begin{figure*}
\begin{center}
	\includegraphics[width=0.8\textwidth]{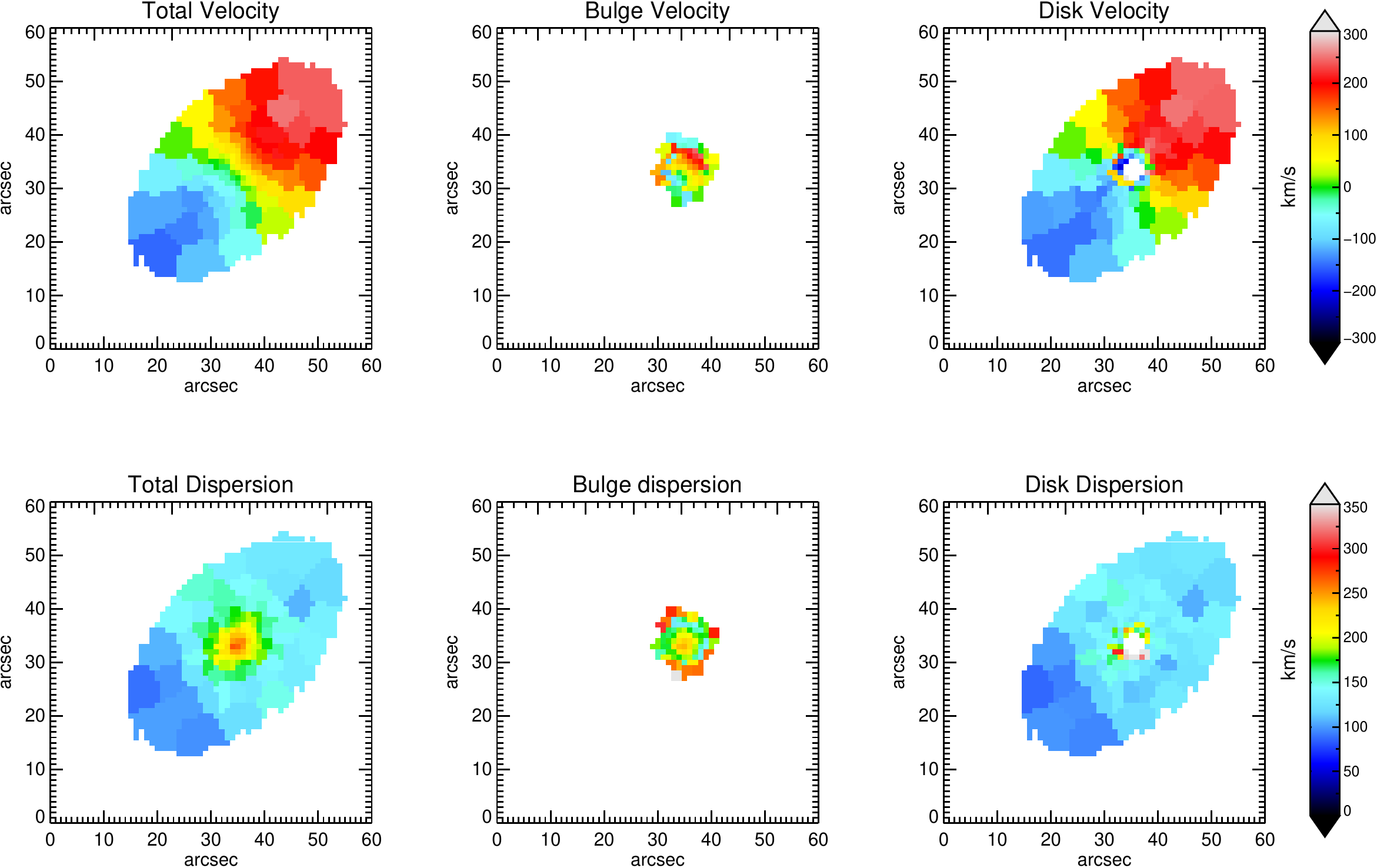} 
 \caption{Kinematic maps for a one component fit (left) and two component fit (centre and right) for NGC~7671.   \label{fig:kin7671}} 
\end{center}
\end{figure*}

\begin{figure*}
\begin{center}
	\includegraphics[width=0.8\textwidth]{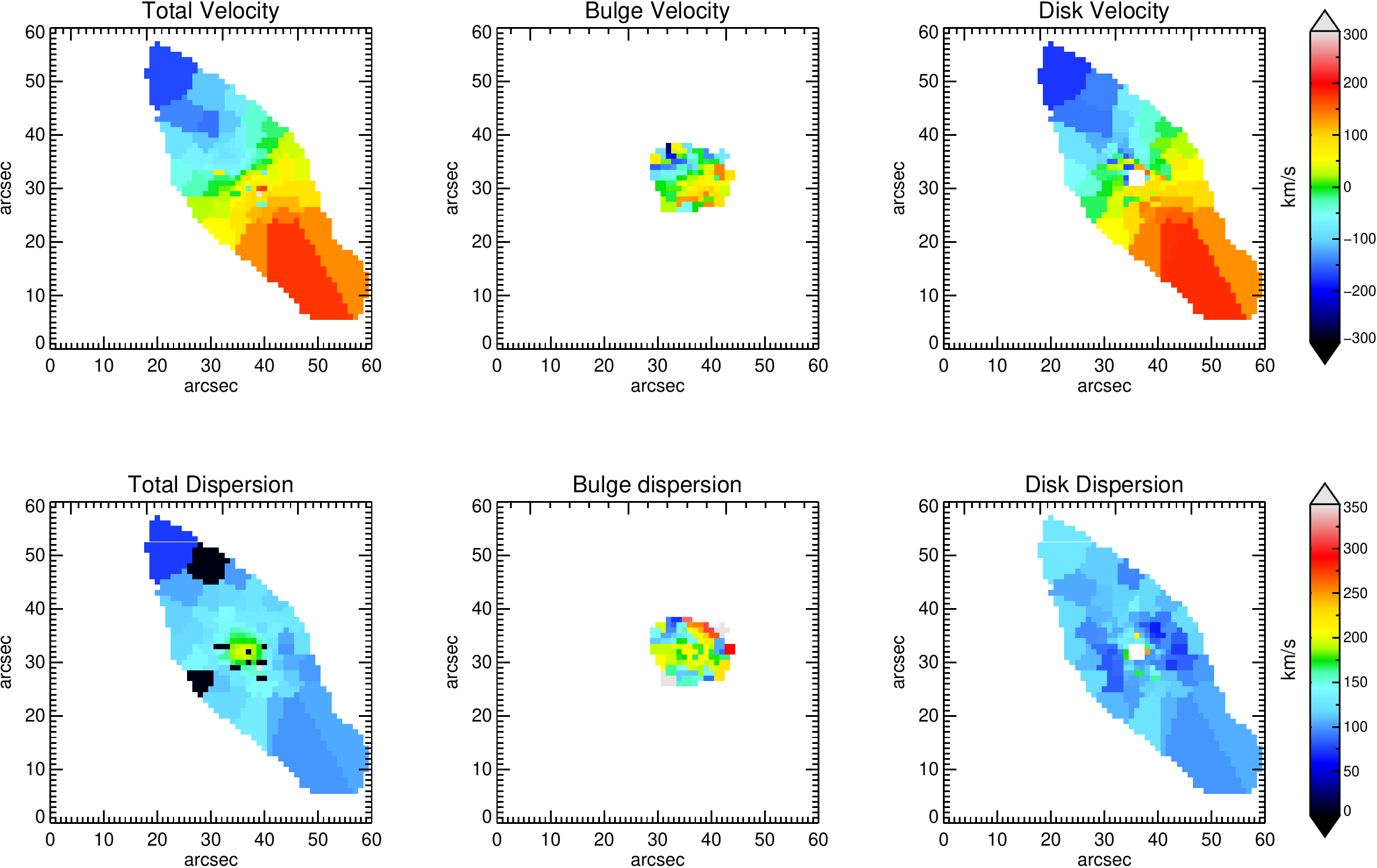} 

 \caption{Kinematic maps for a one component fit (left) and two component fit (centre and right) for NGC~6427. \label{fig:kin6427}} 
\end{center}
\end{figure*}

Figure~\ref{fig:kin7671} and Figure~\ref{fig:kin6427} show the resulting 
decomposed kinematic components. Again, the decompositions result in two 
very coherent components which match the photometric decomposition very well, 
and, as for NGC~528, both show cold rotating discs and hot rotating bulges.
Comparing Figures \ref{fig:bulger}, \ref{fig:kin7671} and \ref{fig:kin6427} 
there is a clear similarity in the kinematic components of these 
three galaxies as well as the photometric components. 
This suggests that not only do photometric decompositions
represent individual stellar components, but the underlying kinematics of 
these components may potentially be deduced from their photometric parameters. 
Of course a much larger, more varied sample of galaxies will be required to 
determine if this relation holds in general.

\subsection{Stellar Populations}

\begin{figure*}
\begin{center}
	\includegraphics[width=0.8\textwidth]{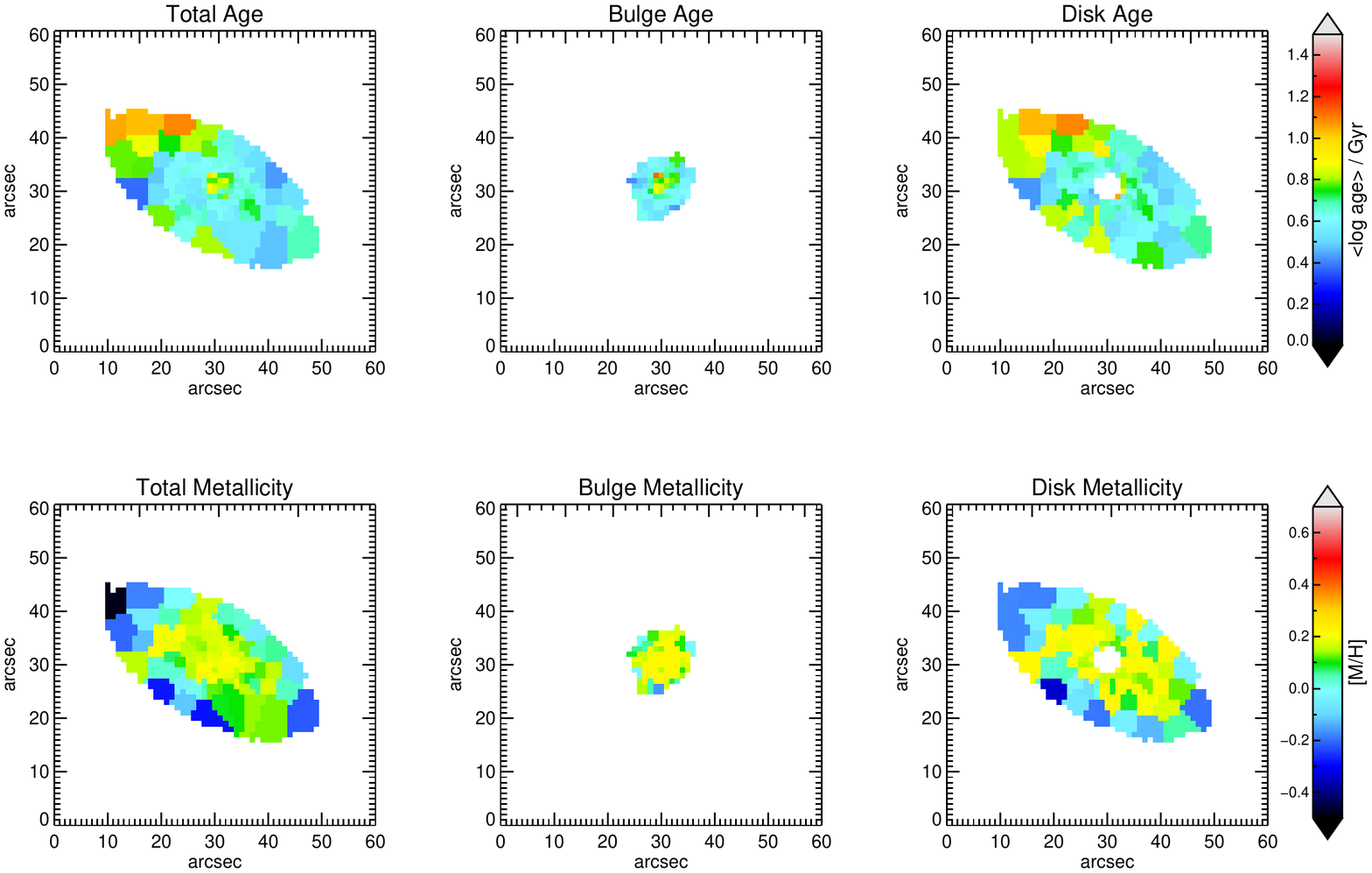} 
        \caption{NGC~528: maps for the luminosity weighted ages and metallicities of the single stellar population spectra selected by 
         pPXF for the one component kinematic model (left) and 
         the two component kinematic
         model (centre and right).
          Bins where the
          flux of the component contributes less than 30\% of the
          total flux are not included. \label{fig:ssps528}}
\end{center}
\end{figure*}

\begin{figure*}
\begin{center}
	\includegraphics[width=0.8\textwidth]{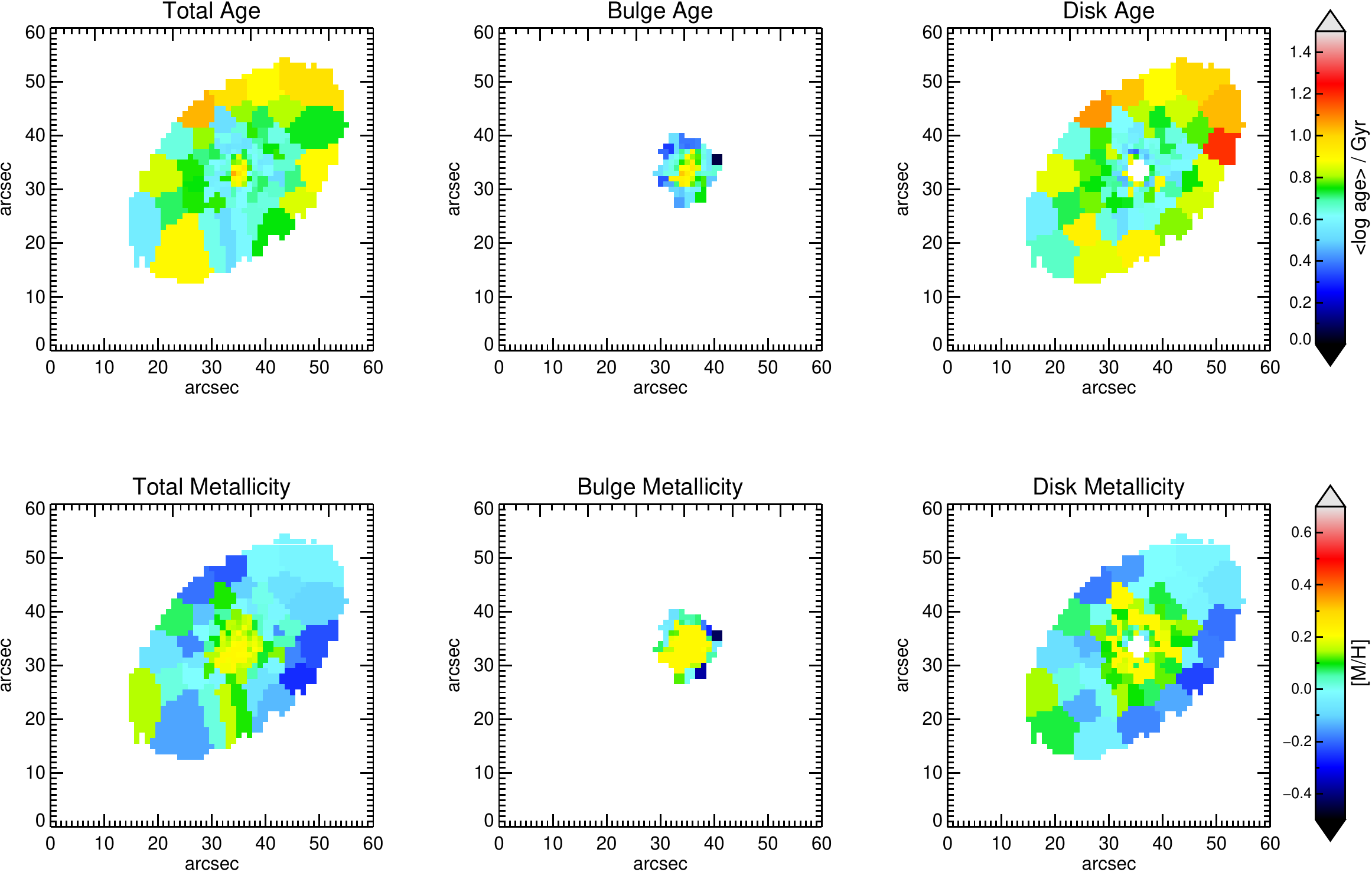} 
        \caption{NGC~7671: maps for the luminosity weighted ages and metallicities of the single stellar population spectra selected by 
         pPXF for the one component kinematic model (left) and 
         the two component kinematic
         model (centre and right).
          Bins where the
          flux of the component contributes less than 30\% of the
          total flux are not included. \label{fig:ssps7671}}
\end{center}
\end{figure*}

\begin{figure*}
\begin{center}
	\includegraphics[width=0.8\textwidth]{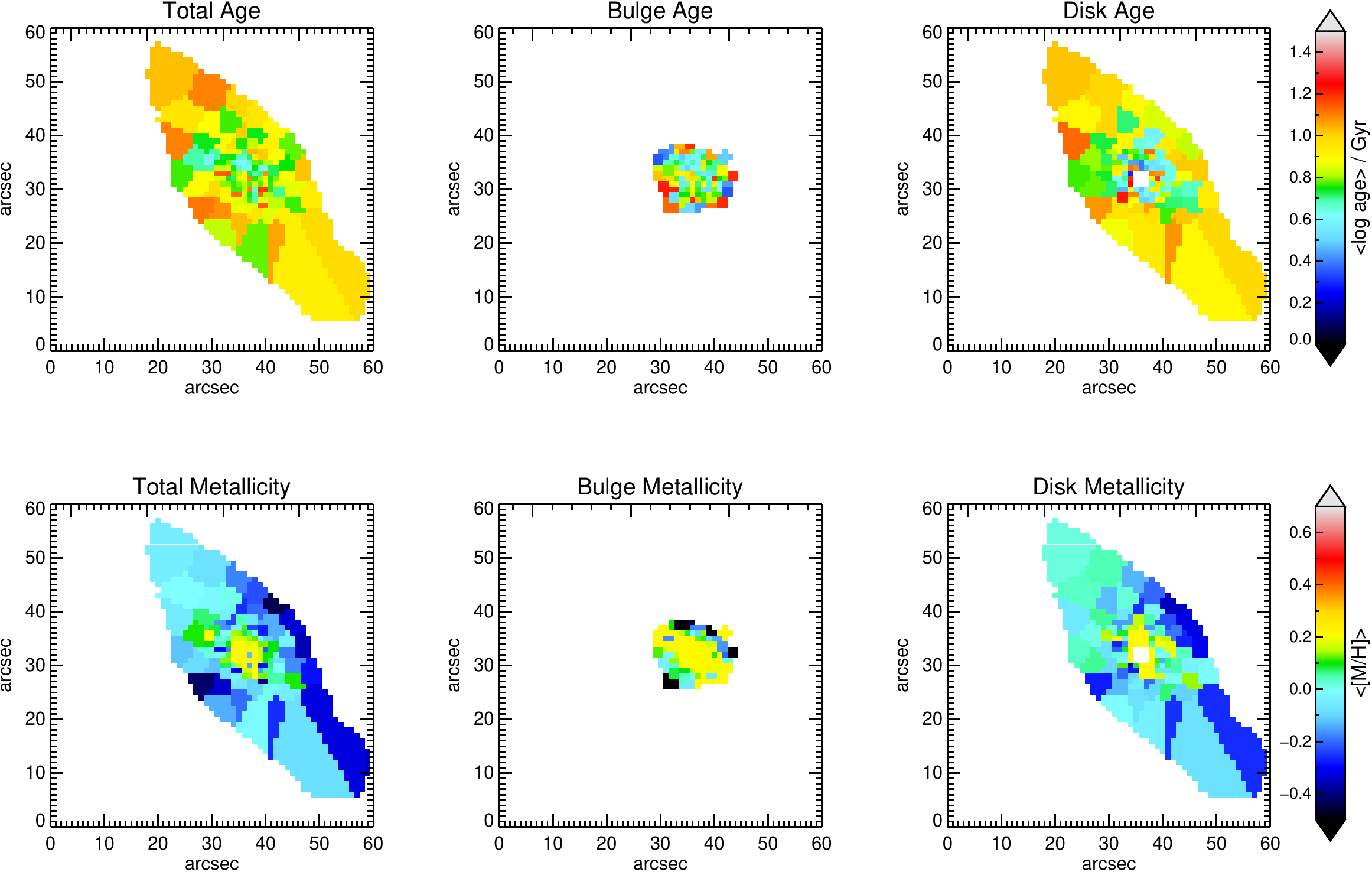} 
        \caption{NGC~6427: maps for the luminosity weighted ages and metallicities of the single stellar population spectra selected by 
         pPXF for the one component kinematic model (left) and 
         the two component kinematic
         model (centre and right).
          Bins where the
          flux of the component contributes less than 30\% of the
          total flux are not included. \label{fig:ssps6427}}
\end{center}
\end{figure*}

Since our fits also left the choice of stellar population completely
free, as well as the weight assigned to each stellar population, they 
provide a measure of the age and metallicity of each
component in each spatial bin, which yields further information on a
galaxy's past history.  In particular, we can go beyond the analysis
undertaken by \citet{Johnston2014} to look for spatial variations in
the properties of the population within a single component.  To
illustrate the potential in such data, Figures~\ref{fig:ssps528}, \ref{fig:ssps7671} and \ref{fig:ssps6427} show
luminosity-weighted log age and metallicity of the single stellar population 
spectra selected by pPXF for the bulge and disc
\begin{equation}
<\textnormal{log age}>_{\rm component}=\frac{\sum (w_{\rm component} \times \textnormal{log age})}{\sum w_{\rm component}},
\end{equation}

\begin{equation}
<[\textnormal{M/H}]>_{\rm component}=\frac{\sum (w_{\rm component} \times \textnormal{[M/H]})}{\sum w_{\rm component}},
\end{equation}
where $w_{\rm component}$ are the weights assigned to the set of spectral templates used to 
fit each component.

Though the age and metellicity maps are more noisy than those of the kinematics, structure in
the populations is clearly visible. The maps display an intriguing variation in populations across the components, 
with the three bulge components appearing younger,
older, and of a similar age relative to their discs, while all three bulges are systematically more metal rich than their surrounding disc components.

It is interesting to compare this to ages and metallicities found for S0s in the Virgo cluster by \citet{Johnston2014}, who showed that
all the bulge components appeared to be younger than the discs. This suggests a difference in the evolutionary processes of
cluster S0 galaxies and the field S0s studied here. 
Intriguingly, at larger radii there are also the
first indications of structure in the metallicity and ages of the
individual components that might allow us to go beyond a monolithic
view of the formation of bulge and disc.

However, we must await the analysis of a larger 
sample to see how prevalent such features seen here may be and what implications they hold 
for the bigger picture of galaxy evolution. We also emphasise that this method thus far is primarily designed for decomposing the kinematics of the components, 
and exploring the most effective method for extracting accurate ages and metallicities is something that 
we intend look at in more detail in the future.

\section{Conclusions}
\label{sec:conc}
In this paper, we have developed a method to find the kinematics and stellar populations 
of the bulge and disk components, when their relative surface brightness 
is constrained to be the one provided by a previous photometric decomposition.  

An initial application of this technique to data cubes for three CALIFA S0
galaxies produces a remarkably coherent picture:
these galaxies really can be separated into a cold rotating disc
component and a hot rotating bulge component that match the
expectations very well. Thus, at least for this test
sample, the bulge and the disc do seem to represent plausible distinct
components. 

For this decomposition to be effective we have shown that the data ideally needs to be
voronoi-binned to at least a signal-to-noise ratio of 25. To ensure that the true best fit
to the data is found, it is best to run over a grid of component velocities for each voronoi bin.
This allows the kinematics to be extracted 
effectively in this case down to a component flux contribution of ~30\% of the total flux;
a level easily low enough to get detailed information on kinematics and stellar populations 
for each component. 


In terms of the kinematics themselves, it appears that all three
bulges are rotating; a finding which, given the low S\'{e}rsic-indices 
found for the bulges in the GALFIT photometric decomposition, 
agrees with previous literature on the disc-like characteristics often 
displayed by flatter bulges. With the ability to separate the kinematics of the 
bulge from those of the disc provided by this method, we can go beyond previous studies
and look at the kinematics purely of the bulge component to see in what way they relate to the photometric parameters 
of the galaxy.
Obtaining the kinematics of the components individually also provides useful raw material in modelling 
each component dynamically, presenting interesting opportunities to explore how the mass is distributed 
in the galaxy, and how this compares to the luminosity distribution.


We have also shown that there is information in the model stellar
populations that produce the best fit. 
Previous studies into 
population gradients in S0 galaxies have produced a wide range of results, finding
older bulge components than discs \citep{Caldwell1983, Bothun1990, Norris2006,Gonzalez2015}, 
similar ages in both components \citep{Peletier1996, Mehlert2003, Chamberlain2011},
and older discs than bulges 
\citep{Fisher1996, Bell2000,  Kuntschner2000, MacArthur2004}. Studies specifically into 
S0 galaxies residing in clusters tend to find predominantly younger bulges
than discs \citep{Bedregal2011,Johnston2012}, indicating that these S0s are most likely the 
evolutionary result of gas stripping from spiral galaxies, in which the ``last gasp'' 
of star formation occurred in their central regions, enhancing the bulge light.

The galaxies in this sample are found in the field and 
reflect this variation in populations gradients found in the studies mentioned above, suggesting 
that these galaxies, in contrast to S0s located in clusters, are not the result of one single evolution scenario.

In addition to these broad population differences, there are intriguing indications of structure within the 
metallicity and age maps of the decomposed bulges and discs, showing 
the potential of the method in revealing more detailed galaxy features. 
Not much should be inferred for such a small sample, but a systematic 
analysis of a larger number of galaxies could tell us a great deal about 
the channels through which these systems form and evolve.
Fortunately, there is now no shortage of IFU
observations of galaxies on which this technique could be used.

Finally, note that although we have concentrated on applying
this method to the decomposition of galaxies into discs and bulges, it
could equally well be applied to any other attempt to separate
galaxies into components, such as multiple nuclei in giant
ellipticals, cores and halos in cD galaxies, and even just
line-of-sight blended images.  It is also worth noting that any
residuals to a fit, whether photometric or spectroscopic,
 are essentially just another component that can be
treated in the same way, so one could, for example, look at the
kinematics and stellar populations of spiral arms that might be left
after the fitting of a smooth disc component to a spiral galaxy.  As a
general method, kinematic spectral decomposition constrained by photometric
decomposition has many applications.

\section*{Acknowledgements}
MC acknowledge support from a Royal Society University Research Fellowship.



\bibliographystyle{mnras}
\bibliography{paper3} 


\bsp	
\label{lastpage}
\end{document}